\documentclass[]{article}

\usepackage{epsfig}

\def\be{{\bar\eta}}
\def\sg{{\sigma}}

\begin{document}


\title{First-principles modeling of strain \\
in perovskite ferroelectric thin films}

\author{
OSWALDO DI\'EGUEZ \\
Department of Materials Science and Engineering, \\
Massachusetts Institute of Technology, \\
77 Massachusetts Avenue, Cambridge MA 02139, USA \\
\\
DAVID VANDERBILT \\
Department of Physics and Astronomy, \\
Rutgers University, \\
136 Frelinghuysen Rd, Piscataway NJ 08854, USA}

\maketitle

\begin{abstract}
We review the role that first-principles calculations
have played in understanding the effects of substrate-imposed
misfit strain on epitaxially grown perovskite ferroelectric films.
We do so by analyzing the case of BaTiO$_3$, complementing our
previous publications on this subject with unpublished data to help
explain in detail how these calculations are done.
We also review similar studies in the literature for other perovskite
ferroelectric-film materials.
\bigskip
\end{abstract}


\section{Introduction}

Ferroelectrics materials are of technological importance due to their ability to
sustain a macroscopic polarization that can be switched by the action of an
electric field.
In particular, the demand for smaller industrial components has drawn attention
in the last few years to ferroelectrics in film form.
These can sometimes be grown epitaxially in a coherent fashion
with thicknesses beyond tens of nanometers.
Part of what makes them interesting is that the properties of the
thin films can differ significantly from those of the corresponding bulk form.
Several factors contribute to these differences, e.g., the thickness of the
film, the electrical boundary conditions imposed on it, the orientation of the
substrate upon which the film grows, the presence of defects or stoichiometric
variations, and the misfit strain imposed by
the substrate on the film. 
The reader interested in the science of ferroelectric films can consult the
recent reviews \cite{Ahn2004,Dawber2005,Junquera2006,Kornev2006,
Posadas2007,Rabe2005,Schwarzkopf2006,Scott2007}, especially the one by
Rabe \cite{Rabe2005} that specifically addresses the theory of epitaxial
strain effects.

In this paper we will be concerned with the role of the misfit strain
in the structural behavior of ferroelectric thin films.
In doing so we will restrict ourselves to perovskites, one of the
most representative and studied families of ferroelectric materials.
We will review how first-principles theories can be used to isolate
the effect of epitaxial strain from the other film-related effects mentioned above.
Section \ref{sec_pertsev} introduces the structural phase diagrams that
have proven to be
very useful in understanding the properties of ferroelectric films.
Section \ref{sec_batio3} explains in detail how first-principles calculations
can be used to obtain those diagrams, focusing on the case of
BaTiO$_3$ as a prototypical example of a perovskite ferroelectric.
This section is based in two of our previous publications
\cite{Dieguez2004,Dieguez2005}, complemented by expanded results
and detailed explanations that have not previously been published.
In Sec.~\ref{sec_others} we review similar studies that have
been carried out for ferroelectric films composed of materials other
than BaTiO$_3$.  Finally, we summarize briefly in Sec.~\ref{sec_summary}.


\section{Pertsev diagrams}
\label{sec_pertsev}

\begin{figure}
\centerline{\epsfxsize=8cm \epsfbox{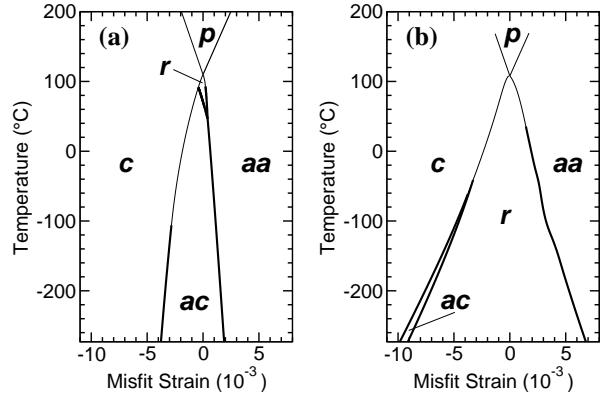}}
\caption{Phase diagrams of epitaxial BaTiO$_3$ as predicted by the theory
         of Pertsev {\em et al.}\protect\cite{Pertsev1998}:
         (a) Using the parameters quoted in
             Ref.~\protect\cite{Pertsev1998}. 
         (b) Using the parameters quoted in
             Ref.~\protect\cite{Pertsev1999}.
         First- and second-order phase transitions are represented by
         thick and thin lines, respectively.
         (Reprinted from Ref.~\protect\cite{Dieguez2004}).}
\label{fig_diagramsPERTSEV}
\end{figure}

In a seminal paper, Pertsev, Zembilgotov and Tagantsev \cite{Pertsev1998}
introduced the concept of mapping the equilibrium structure of a ferroelectric
perovskite material as a function of temperature and misfit strain, thus producing a
``Pertsev phase diagram'' of the observable epitaxial phases.  
The effect of epitaxial strain is isolated from other aspects of thin-film
geometry by computing the structure of the {\em bulk} material with homogeneous
strain tensor constrained to match a substrate having square surface symmetry
and a given in-plane lattice constant.
In addition, short-circuit electrical boundary conditions are imposed,
equivalent to ideal electrodes above and beneath the film
\cite{Pertsev1998}.
Examples of such diagrams are shown in Fig.~\ref{fig_diagramsPERTSEV};
the labeling of the phases therein is clarified in Table~\ref{tab_phases}.
Given the recognized importance of strain in determining the properties of 
thin-film ferroelectrics, Pertsev diagrams have proven to be of enormous
value to experimentalists seeking to interpret the behaviour of
epitaxial thin films and heterostructures.

\begin{table}[b]
\caption{Summary of possible epitaxial BaTiO$_{3}$ phases.
         In-plane cell vectors are fixed at ${\bf a}_1 = a \hat x$,
         ${\bf a}_2 = a \hat y$.  Columns list, respectively:
         phase label; space group; out-of-plane lattice vector; number of
         free internal displacement coordinates; and form of the
         polarization vector.}
\centering
\begin{tabular}{@{}ccccc}
\hline
 Phase    & SG      & ${\bf a}_3$                             & $N_{\rm{p}}$ 
                                                               & Polarization \\
\hline
 {\em p}  & $P4mmm$ & $c \hat z$                              & 0          
                                                                          & 0 \\
 {\em c}  & $P4mm$  & $c \hat z$                              & 3          
                                                                & $P_z\hat z$ \\
 {\em aa} & $Amm2$  & $c \hat z$                              & 4         
                                                         & $P(\hat x+\hat y)$ \\
 {\em a}  & $Pmm2$  & $c \hat z$                              & 4          
                                                                  & $P\hat x$ \\
 {\em ac} & $Pm$    & $c_\alpha \hat x + c \hat z$            & 8          
                                                    & $P \hat x + P_z \hat z$ \\
 {\em r}  & $Cm$    & $c_\alpha (\hat x + \hat y) + c \hat z$ & 7          
                                          & $P (\hat x+\hat y ) + P_z \hat z$ \\
\hline
\end{tabular}
\label{tab_phases}
\end{table}

In Ref.~\protect\cite{Pertsev1998}, the mapping was carried out using
a phenomenological Landau-Devonshire model taken from the literature.
Within this kind of theory, a thermodynamical potential describing
the behaviour of the material is expanded in the relevant degrees of freedom.
In the case of Ref.~\protect\cite{Pertsev1998},
the basic degrees of freedom were stress and
polarization, with the temperature being incorporated via a linear
temperature dependence of certain expansion parameters.
For fixed misfit strain and temperature, a minimization over the three
components of the polarization was carried out. 
Depending on the direction the polarization points at after the optimization,
the film is in a distinctive phase labeled as indicated in
Table~\ref{tab_phases}.

This approach should give excellent results in the regime of temperature
and strain in which the model parameters were fitted.
However, it will generally be less accurate when extrapolated to other regimes.
As an illustration of this, we pointed out in Ref.~\cite{Dieguez2004}
that different sets of parameters can lead to significantly different 
diagrams even when exactly the same model is used, as shown
in Fig.~\ref{fig_diagramsPERTSEV}.
Of course, such an empirical approach is also limited to materials
for which all the needed experimental information is available.


\section{First-principles calculations: the example of BaTiO$_3$ films}
\label{sec_batio3}

In this section we describe in detail how to apply a first-principles theory
to the study of misfit strain effects in single-domain perovskite-oxide thin
films grown on a substrate with square-lattice symmetry. 
In particular, we focus on the prototypical example of barium titanate.
In Sec.~\ref{sec:full} we show how this is done at a microscopic level,
considering the physics of the individual electrons and ions in the film;
this is what we call full first-principles calculations.
In Sec.~\ref{sec:ld} we show how to apply a Landau-Devonshire theory in
which first-principles calculations are used to find the coefficients in a
Taylor expansion of the thermodynamical potential that describes the 
behaviour of the film; this is what we call first-principles-based
calculations.
In both cases, the calculations are done for zero temperature.
Finally, in Sec.~\ref{sec:heff} we show the results of applying a
refined first-principles-based theory to take into account the effects of
finite temperature.


\subsection{Full first-principles calculations}
\label{sec:full}

\subsubsection{Theoretical details}

First-principles (i.e., {\em ab initio}) theories
make use of the fundamental laws that govern the behaviour of electrons and
nucleii to derive the macroscopical properties of a material.
Among these, density-functional theory (DFT) \cite{Hohenberg1964} as implemented
in the Kohn-Sham approach \cite{Kohn1965} is the most widely used to
deal with systems such as those described here, since relatively large
systems with many electrons can be treated accurately with a modest
expenditure of computer time.
Nowadays, several sophisticated computer codes capable of running such
DFT calculations are available as open-source or inexpensive packages,
including, e.g.,
{\sf ABINIT}\cite{abinit}, the {\sf Quantum-ESPRESSO}\cite{espresso} suite of
programs, {\sf Siesta}\cite{siesta}, and {\sf VASP}\cite{vasp}.

The first-principles DFT calculations described in this subsection
were carried out using the {\sf VASP}\cite{vasp} software package.
The electron-ion interaction was described by the projector augmented-wave
method \cite{Blochl1994}; semicore
electrons are included in the case of Ba ($ 5s^2 5p^6 6s^2 $) and
Ti ($ 3s^2 3p^6 4s^2 3d^2 $). 
The calculations employed the Ceperley-Alder \cite{Ceperley1980} form of
the local-density approximation (LDA) exchange-correlation functional;
other flavours of this functional, like the generalized gradient aproximation,
do not lead to substantial improvements in the case of BaTiO$_3$ 
\cite{Singh1997}.
We use a plane-wave basis set with a 700\,eV kinetic-energy cutoff and
a $6 \times 6 \times 6$ Monkhorst-Pack \cite{Monkhorst1979}
sampling of the Brillouin zone.

Readers wishing to become more familiar with the terminology used in the
previous two paragraphs, or to know more about the conceptual
framework behind DFT, can consult a recent review on the subject by
Capelle \cite{Capelle2006}; see also references therein.

\subsubsection{Search for the most stable phases}

Given the constituents of the unit cell, DFT codes allow one to carry out
a search for the structural configuration with the lowest energy.
We did this for the case of BaTiO$_3$ in Ref.~\cite{Dieguez2004} where,
following Pertsev {\em et al.} \cite{Pertsev1998}, we worked with a bulk-like
periodic system in which the film boundary conditions imposted by the misfit
strain have been applied.
We emphasize that there are no surfaces or interfaces in our calculation.
Essentially, we take a single unit cell from the deep interior of the
film and replicate it periodically in all three dimensions; this periodic
structure is the one used for our calculations.

We performed optimizations of distorted configurations of the
five-atom unit cell of the cubic perovskite structure in order to look for the
six possible phases mentioned in Table \ref{tab_phases}.
Each starting structure had the Ba atoms fixed as determined by the substrate,
and we displaced the Ti and O atoms along the direction of the polarization
vector of each phase.
Then, while retaining the symmetry determined by these displacements,
we relaxed the atomic positions and the out-of-plane cell vector until
the value of the Hellmann-Feynman forces and $zz$, $yz$ and $zx$ stress tensor
components fell below 0.001\,eV/\AA~and 0.005\,eV, respectively.
This process allowed us to determine the lowest-energy structure for each of
the six possible phases.

\begin{figure}
\centerline{\epsfxsize=8cm \epsfbox{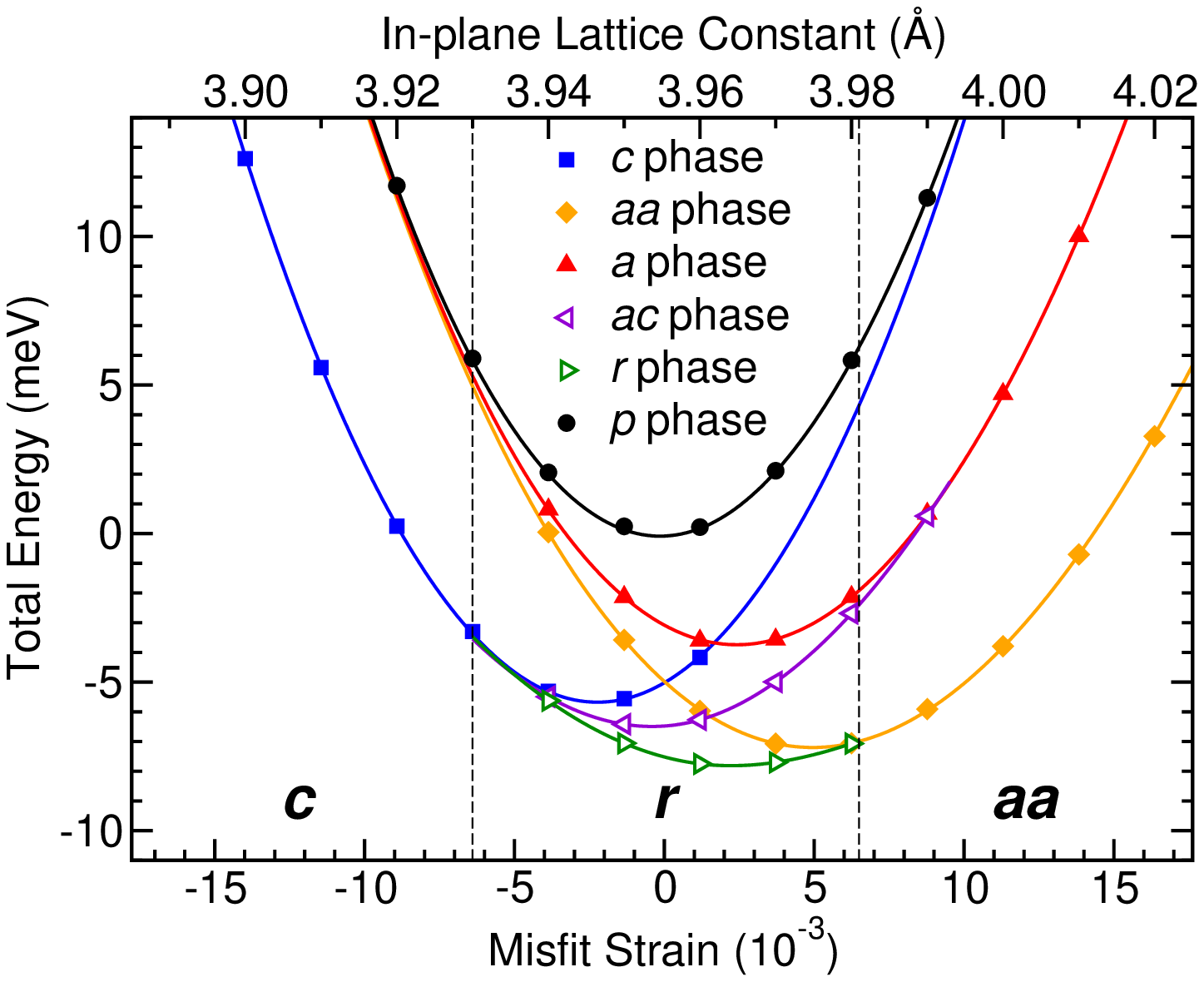}}
\centerline{\epsfxsize=8cm \epsfbox{fig2b.ps}}
\centerline{\epsfxsize=8cm \epsfbox{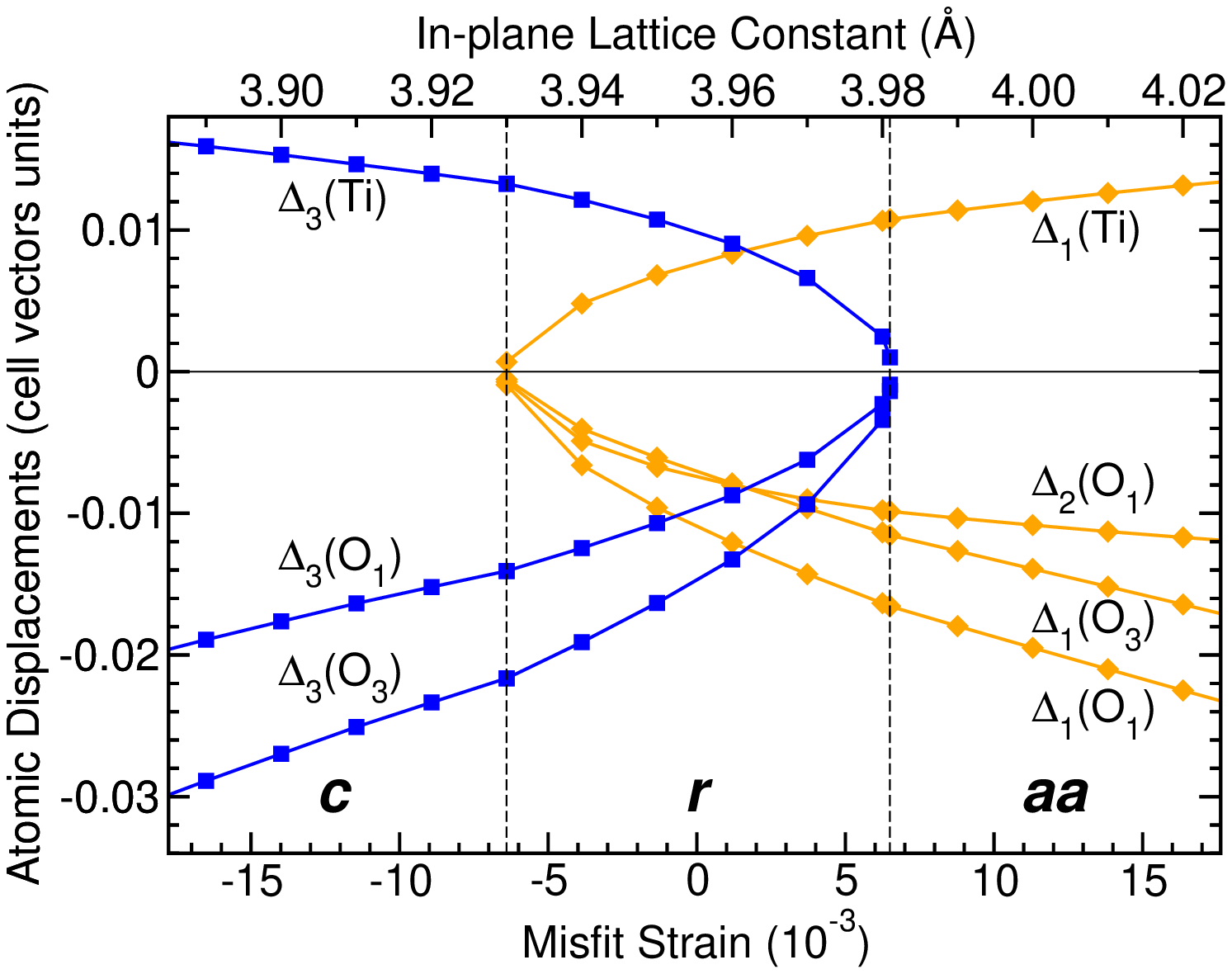}}
\caption{Top panel: energies of the possible epitaxial BaTiO$_3$ phases for
         different misfit strains, as obtained from the full first-principles
         calculations. 
         Center panel: value of the lowest non-trivial eigenvalue of the
         force-constant matrix for the three relevant phases; when it is zero,
         the second-order phase transitions denoted by vertical lines occur.
         Bottom panel: displacements of the atoms from the cubic perovskite
         cell positions,
         for the most energetically favorable configuration at a given misfit 
         strain;
         $\Delta_3\rm{(Ti)}$ labels the displacement of the Ti atom in units
         of the third lattice vector, etc.}
\label{fig_energiesDFT}
\end{figure}

Following this recipe for different values of the in-plane strain on the film, we
obtained the energy curves of Fig.~\ref{fig_energiesDFT} (top panel).
In some cases we started with a structure with a given symmetry, but found
that the system relaxed to a structure with a higher symmetry;
for example, starting with a structure having the symmetry of the $r$
phase at misfit strains exceeding $\sim0.6$\%, we find that it converges
to a structure of $aa$ symmetry (collapse of the green curve onto the
orange curve in the figure).
It can be seen that the $c$ phase is the most stable structure for large
negative (compressive) misfit strains, then the $r$ phase is most stable
for an intermediate strain range, and finally the $aa$ phase is most stable
at large positive (expansive) strains.

Because the phase transitions from $c\rightarrow r\rightarrow aa$ are
both of second order, the curves join each other so smoothly that it
is difficult to locate the precise phase boundaries from the energy
curves alone.  The boundaries can be located much more precisely by
using a stability analysis.  At each value of misfit in the {\em c}
phase, for example, we carry out finite-difference calculations of
$x$ forces and the $xz$ stress as the atomic $x$ coordinates and the
$xz$ strain are varied.  The lowest eigenvalue of the resulting
$6 \times 6$ Hessian matrix is plotted as the blue curve in the
middle panel of Fig.~\ref{fig_energiesDFT}, and its
zero crossing identifies the critical misfit.
A similar analysis is used to consider $z$
displacements and shear strains in the {\em a} and {\em aa} phases
(red and orange lines respectively).
By properly considering the zone-center phonons, the elastic
shear, and linear cross-coupling between these two kinds of degrees
of freedom, this analysis allows
us to locate the second-order phase boundaries much more precisely
than is possible through direct comparison of total energies.
Since the $a$ phase is never the lowest-energy structure for any
misfit strain (see top panel), we can ignore it, and we locate the
$c\leftrightarrow r$ and $r\leftrightarrow aa$ transitions as
indicated by the vertical dashed lines in the figure.

With first-principles calculations it is also possible to inspect the
individual displacements of the atoms in each structure, as shown in 
the bottom panel of Fig.\ref{fig_energiesDFT}.
At the phase transition points, displacement patterns arise that characterize
the phases on both sides of the transition.

The three panels in Fig.~\ref{fig_energiesDFT} give therefore a consistent
picture showing how the structural properties of the film vary as a function
of the misfit strain imposed on it.  For large compressive strains, the
lowest energy corresponds to the {\em c}
phase; on the other hand, for large tensile strains the {\em aa} phase is
favored.
At a misfit strain of $s_{\rm max}(c) = -6.4 \times 10^{-3}$ ($a = 3.930$\,\AA),
there is a second-order transition
from the {\em c} phase to the {\em r} phase, with the polarization in the 
{\em r} phase continuously rotating away from the {\em z} direction as the
misfit strain increases. 
At misfit strain $s_{\rm min}(aa) = 6.5 \times 10^{-3}$ ($a = 3.981$\,\AA), the
{\em r} phase polarization completes
its rotation into the {\em xy} plane, resulting in another second-order 
transition to the {\em aa} phase.

\subsubsection{Comments on the underlying assumptions}

\begin{figure}
\centerline{\epsfxsize=8cm \epsfbox{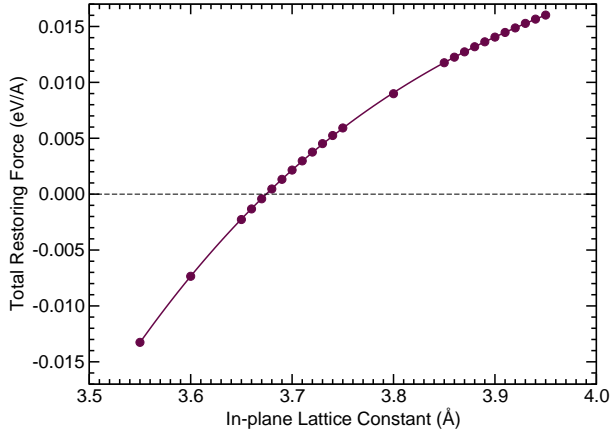}}
\caption{Value of the restoring force on the oxygen octahedra when they
are subjected to a rotation about the [001] axis of $0.23^\circ$
in a pattern of M$_3$ symmetry.  A negative value indicates an
instability leading to a doubled unit cell with oppositely rotated
octahedra, but this only happens at a very high compressive strain
(the equilibrium lattice constant is 3.96\,\AA).}
\label{fig_rotations}
\end{figure}

Here we comment briefly on the effects of the assumptions made in the construction
of this first-principles Pertsev diagram. 
In principle, we should consider the possibility of equilibrium structures with
larger unit cells, particularly those with cell-doubling octahedral rotations
(i.e., tilts).  Such rotations
have been shown to be important in SrTiO$_3$, and could condense in
BaTiO$_3$ under sufficiently large misfit strains.
As an example, we have checked that the paraelectric phase of the film is
stable with respect to octahedral rotations about the [001] direction (with
M$_3$ symmetry) up to an epitaxial compressive strain of 
$-70.9 \times 10^{-3}$ ($a = 3.675$\,\AA), far larger than those likely to be 
experimentally relevant (see Fig.\ \ref{fig_rotations}).

In addition, while we have studied only the effects of epitaxial strain,
other physical effects may also be relevant to the structure and properties of
thin films, such as atomic rearrangements at the film-substrate interface and
free surface and the instability to the formation of multiple domain
structures \cite{Li2003}.
Finally, our theory relies on the LDA to compute the exchange and correlation
terms in DFT.  This introduces small systematic errors in the
calculation, the most important of which is probably the error in
the equilibrium lattice constant.  However, such errors are well
understood and well characterized in perovskites, and tend to be
similar for different materials of this class, so that there is a
tendency for cancellation of errors when making relative comparisons
of quantities such as misfit strains (see, for example, Ref.~\cite{KingSmith1994}).

It should be kept in mind, however, that for both films and superlattices, the
assumptions that the system is in a single domain and that the epitaxial 
strain strongly dominates other factors will not be valid in all cases.
Phase diagrams including multiple-domain states have, for example, been 
discussed in Refs.~\cite{Speck1994,Pertsev2000,Li2003}. 
Other influences that may be important include surface relaxation and
reconstruction, atomic and electronic rearrangements at the
interface, imperfectly compensated macroscopic electric fields,
deviations from stoichiometry, and the presence of defects.


\subsection{First-principles-based calculations: a Landau-Devonshire model}
\label{sec:ld}

In a Landau-Devonshire approach, a ferroelectric system is described
by a thermodynamical potential that is expanded as a Taylor series of
the relevant degrees of freedom.
For example, in the work of Pertsev and collaborators \cite{Pertsev1998}, the
model is built by starting with the bulk free energy expanded in polarization
and stress.
The reference state is the paraelectric cubic perovskite phase at the bulk 
critical temperature $T_{\rm c}$, and the parameters are fit to reproduce 
experimental observations of the behavior near the bulk ferroelectric
transition. 
For the dependence on epitaxial strain, a Legendre transformation is then made to 
obtain the potential as a function of polarization and misfit strain. 
Because of the way in which the parameters are fit, this Landau-Devonshire
potential will give its most accurate results for small misfit strains and
temperatures near the bulk $T_{\rm c}$.

It is also possible to develop models in the same spirit, but constructed from
first-principles calculations.
Compared with the phenomenological approach, such a first-principles-based theory
has the advantage that the information needed for the model fit can readily
be calculated in a consistent and accurate manner within the framework of the
chosen first-principles approach.  Then, once the fitting has been completed,
such models can be used to compute detailed properties orders of magnitude
faster than could be done using the full first-principles calculations.
Thus, they can be applied to larger systems where the full
calculations would be impractical.
In this Section, we summarize how one such model \cite{Dieguez2005} 
can be applied to study BaTiO$_3$, providing additional details that were
not included in the previous publication in order to clarify
how the expansion coefficients were obtained.

\subsubsection{Formalism}

The starting point of this analysis is the parameterized
total-energy expression presented by King-Smith and Vanderbilt in
Ref.~\cite{KingSmith1994}.  This is a Taylor expansion
around the cubic perovskite structure in terms of the six independent
components $\eta_i$ of the strain tensor ($i$ is a Voigt index,
$i=$1-6) and the three Cartesian components $u_{\alpha}$
(${\alpha} = x,y,z$) describing the amplitude of the soft mode
defined by the pattern of eigen-displacements associated with the
smallest eigenvalue of the (zone-center) force-constant matrix.

\begin{figure}
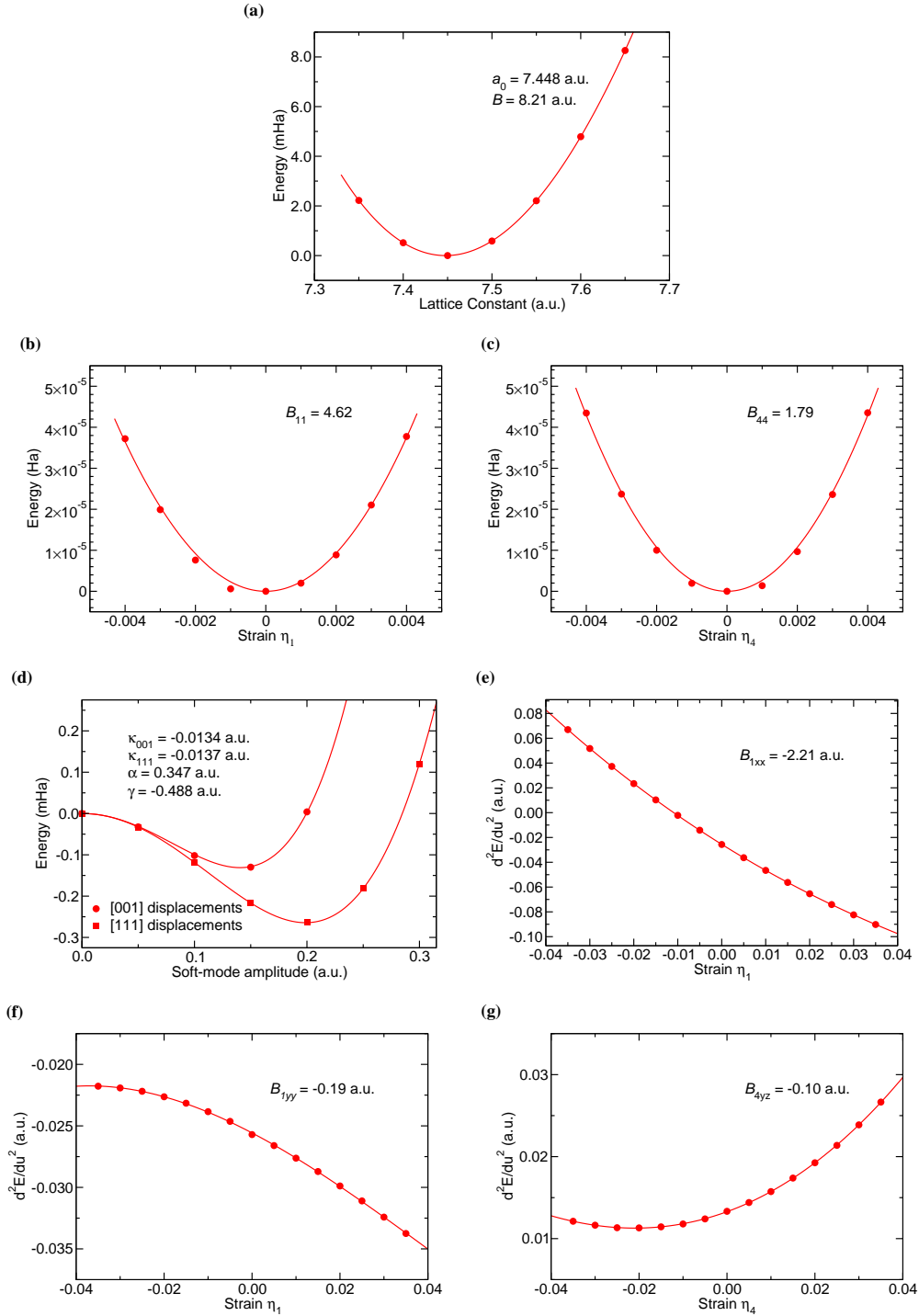

\centerline{\epsfysize=4.5cm \epsfbox{fig4a.eps}}
\vspace{3mm}
\centerline{\epsfysize=4.5cm \epsfbox{fig4b.eps}
            \hspace{3mm}
            \epsfysize=4.5cm \epsfbox{fig4c.eps}}
\vspace{3mm}
\centerline{\epsfysize=4.5cm \epsfbox{fig4d.eps}
            \hspace{3mm}
            \epsfysize=4.5cm \epsfbox{fig4e.eps}}
\vspace{3mm}
\centerline{\epsfysize=4.5cm \epsfbox{fig4f.eps}
            \hspace{3mm}
            \epsfysize=4.5cm \epsfbox{fig4g.eps}}
\vspace{5mm}
\caption{Results of first-principles calculations used to fit the coefficients of the
first-principles-based model for bulk BaTiO$_3$ as described in this section.
In each panel, each symbol corresponds to a full first principles calculation, while
each curve is a low-order polynomial fit to the data.
The ground-state lattice constant and bulk modulus are obtained from the
data in (a), through the use of the Birch equation.
$B_{11}$ and $B_{44}$ are related to the curvature of the third-order polynomials
of graphs (b) and (c); $B_{12}$ is obtained from the relation
$B=B_{11}+2B_{12}$.
The coefficient $\kappa$ is half the lowest eigenvalue of the
force constant matrix, obtained using finite differences and the ability of
the first-principles programs to compute forces on the atoms.
The coefficients $\alpha$ and $\gamma$ are obtained from the double-well
sixth-order polynomial in (d).
Finally, $B_{1xx}$, $B_{1yy}$, and $B_{4yz}$ are proportional to the first
derivative at zero strain of the third-order polynomials of (e), (f), and (g).}
\label{fig_coeff}
\end{figure}

The energy (per unit cell) of the bulk material can be expressed as a sum
    \begin{equation}
    E = E^{\rm elas}(\{\eta_i\}) + E^{\rm soft}(\{u_{\alpha}\})
      + E^{\rm int}(\{\eta_i\},\{u_{\alpha}\}) ,
    \label{eq:energy}
    \end{equation}
of a term $E^{\rm elas}$ arising purely from strain, a term $E^{\rm soft}$ arising
purely from the soft-mode amplitude, and a term $ E^{\rm int}$ describing the
interactions between these two kinds of degrees of freedom.
The zero of energy is taken to correspond to the cubic reference structure.
For crystals with cubic symmetry, the strain energy is given,
to second order in strain, by
    \begin{eqnarray}
    E^{\rm elas}(\{\eta_i\})
      & = & \frac{1}{2} B_{11}
            ( \eta_1^2 + \eta_2^2 + \eta_3^2 )
            + B_{12}
            ( \eta_1 \eta_2 + \eta_2 \eta_3
            + \eta_3 \eta_1 )
            + \frac{1}{2} B_{44}
            ( \eta_4^2 + \eta_5^2 + \eta_6^2 ) ,
    \label{eq:elas}
    \end{eqnarray}
where $B_{11}$, $B_{12}$, and $B_{44}$ are related to the elastic
constants of the crystal by factors of the cell volume.  The
soft-mode energy given in Ref. \cite{KingSmith1994}
contains terms up to fourth-order in the soft-mode amplitude,
    \begin{equation}
    E^{\rm soft}(\{u_{\alpha}\})
      = \kappa u^2 + \alpha u^4 + \gamma ( u_x^2 u_y^2 + u_y^2 u_z^2
                                          + u_z^2 u_x^2 ) ,
    \label{eq:soft}
    \end{equation}
where $u^2 = u_x^2 + u_y^2 + u_z^2$, $\kappa$ is twice the
soft-mode eigenvalue, and $\alpha$ and $\gamma$ are the two
independent symmetry-allowed fourth-order coefficients describing the
cubic anisotropy.  Finally, the interaction between the strains and
the soft-mode amplitude is given to leading order by
    \begin{eqnarray}
    E^{\rm int}(\{\eta_i\},\{u_{\alpha}\})
      & = & \frac{1}{2} B_{1xx} ( \eta_1 u_x^2 + \eta_2 u_y^2 + \eta_3 u_z^2 )
       + \frac{1}{2} B_{1yy} [ \eta_1 ( u_y^2 + u_z^2 )
                                  + \eta_2 ( u_z^2 + u_x^2 )
                                  + \eta_3 ( u_x^2 + u_y^2 ) ]
          \nonumber \\
      &   & \qquad
       + B_{4yz} ( \eta_4 u_y u_z + \eta_5 u_z u_x + \eta_6 u_x u_y ) ,
    \label{eq:int}
    \end{eqnarray}
where $B_{1xx}$, $B_{1yy}$, and $B_{4yz}$ are the phonon-strain
interaction coefficients. All the coefficients in these three parts
of the total-energy expression can be obtained from
first-principles calculations on a series of distorted structures
as described in Refs.~\cite{KingSmith1994,Dieguez2005}.  In
Fig.~\ref{fig_coeff} we present previously unpublished results of our
first-principles calculations that were used as the basis for fitting
these coefficients for the case of BaTiO$_3$.

Once we have an expression for the energy of bulk BaTiO$_3$, we can 
develop a potential appropriate for describing films.
In the case of coherent epitaxy, where strain elements $\eta_1$, $\eta_2$
and $\eta_6$ are constrained while the others are free, this expression is
    \begin{eqnarray}
    \tilde G & = & (A_{\be\be} \be^2 + A_{\be\sg} \be \sg + A_{\sg \sg} \sg^2)
          + ( B_{\be} \be + B_{\sg} \sg + B ) u_{xy}^2
          + ( C_{\be} \be + C_{\sg} \sg + C ) u_{z}^2
    \nonumber \\
    &   & \qquad + D u_{xy}^4 + E u_{z}^4 + F u_{xy}^2 u_{z}^2
          + H u_{xy}^4 \sin^2\theta \cos^2\theta .
    \label{eq_gtilde}
    \end{eqnarray}
where $\sigma=\sigma_{zz}$ is a possible external uniaxial stress applied
perpendicular to the film, $u_{xy}=(u_x^2+u_y^2)^{1/2}$, and
$\theta=\tan^{-1}(u_x/u_y)$, so that
    \begin{eqnarray}
    u_x & = & u_{xy} \cos\theta , \\
    u_y & = & u_{xy} \sin\theta .
    \end{eqnarray}
The detailed expressions relating the coefficients appearing in
Eq.~(\ref{eq_gtilde}) back to the KSV coefficients in
Eqs.~(\ref{eq:elas}-\ref{eq:int}) are given in Ref.~\cite{Dieguez2005}.

Once we have determined the set of coefficients in
Eq.~(\ref{eq_gtilde}), we can predict the phase diagram as a
function of misfit strain $\be$ and the normal external stress
$\sigma$ by minimizing $\tilde G$ to find
the values of the ground-state soft-mode amplitude components.  For
a fourth-order theory like the present KSV expression, the entire
optimization process can be done analytically, since it is possible
to compute first and second derivatives of $\tilde{G}$ and to do a
stability analysis of the various possible phases.

\subsubsection{Testing the model: zero perpendicular external stress}

\begin{figure}
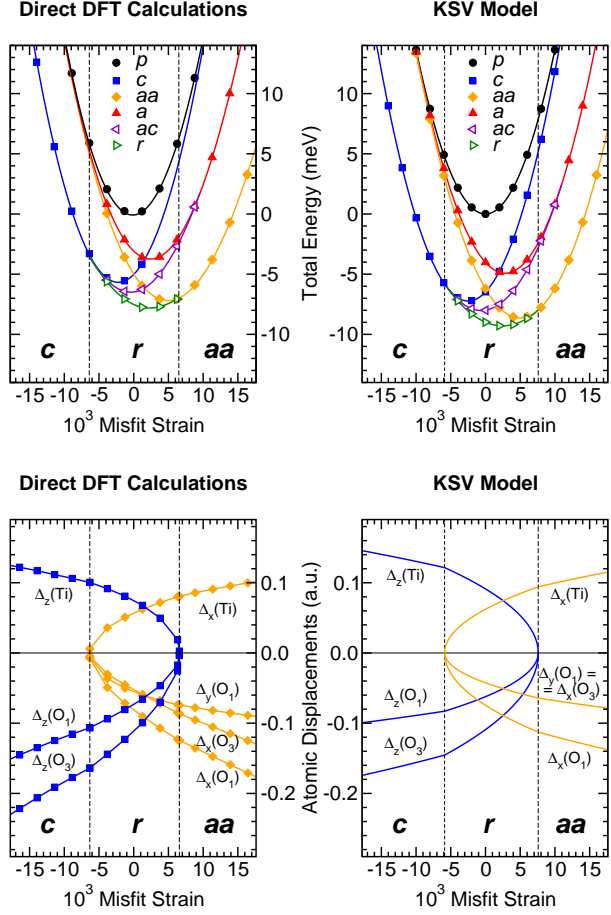

\centerline{\epsfxsize=8cm \epsfbox{fig5a.eps}}
\vspace{5mm}
\centerline{\epsfxsize=8cm \epsfbox{fig5b.eps}}
\caption{Top panel:
         comparison of BaTiO$_3$ energy curves for the six epitaxial phases
	 studied, as obtained from direct DFT
	 calculations\protect\cite{Dieguez2004} (left), and from
	 the KSV theory (right). Energies are relative to the
	 paraelectric structure at zero misfit strain.  The lines
	 in the left panel and the symbols in the right panel are
	 provided as guides to the eye.
         Bottom panel:
         Comparison of BaTiO$_3$ atomic displacements for the most
	 stable phase at each given value of strain, as obtained
	 from direct DFT calculations\protect\cite{Dieguez2004}
	 (left), and from our KSV model (right). 
	 $\Delta_z({\rm
	 Ti})$ indicates the displacement of the Ti atom along the
	 $z$ direction, etc.  Symmetry implies that
         $\Delta_y({\rm Ti})=\Delta_x({\rm Ti})$,
         $\Delta_x({\rm O}_2)=\Delta_y({\rm O}_1)$,
         $\Delta_y({\rm O}_2)=\Delta_x({\rm O}_1)$,
         $\Delta_z({\rm O}_2)=\Delta_z({\rm O}_1)$, and
         $\Delta_y({\rm O}_3)=\Delta_x({\rm O}_3)$.
         The lines in the left panel are guides to the eye.}
\label{fig_btdisp}
\end{figure}

We first consider the usual case in which the external perpendicular stress
$\sigma$ vanishes.
Figure \ref{fig_btdisp} (top) shows the
energy curves of the various phases as predicted by the first-principles-based
KSV theory
(right panel) compared with the full DFT results (left
panel).  The agreement between the two sets of results is very
good, with the small differences arising from two sources.
First, the first-principles calculations in Ref.~\cite{Dieguez2004} (and in
Sec.\ \ref{sec:full}
were performed using the
projector-augmented wave method,\cite{Blochl1994} while the
first-principles calculations used to obtain the KSV coefficients
referred here were performed using ultrasoft pseudopotentials
\cite{Vanderbilt1990}.  Second, there are the intrinsic errors
associated with the use of a Taylor expansion that has been truncated
as described in the previous section; these errors are expected to grow
as the strain and soft mode amplitudes increase.
The bottom panels in Fig.~\ref{fig_btdisp} show the displacements of the atoms
from their centrosymmetric perovskite positions as the strain is varied.
Again, the agreement between the KSV results (right) and the full
DFT results (left) is very good.  In particular, the square-root
behavior predicted by the KSV theory is exhibited by the more
exact DFT calculations.

\subsubsection{Other results}

In Ref.~\cite{Dieguez2005} we studied the effect of external stress $\sigma$
applied perpendicular to the film in addition to the effect of misfit strain.
We did this for BaTiO$_3$ and other seven perovskites, constructing a
phase diagram for each in the two-dimensional space of normal stress
and misfit strain.   We found that the stress-strain phase diagrams obtained
for all of the perovskites show a universal topology with straight-line
phase boundaries meeting at a single crossing point.
A detailed interpretation of the those results was given in
Ref.~\cite{Dieguez2005}, together with an analysis of the variables that drive
each kind of behavior seen for each film.
In the same article we also presented the computed polarization of the films and
related these results to the concept of polarization matching.


\subsection{Finite-temperature calculations}
\label{sec:heff}

The calculations presented so far were all done at zero temperature.
In Ref.~\cite{Dieguez2004}, we extended our study of epitaxial BaTiO$_3$ to
finite temperatures by using
the effective Hamiltonian approach of Zhong, Vanderbilt, and
Rabe.\cite{Zhong1994,Zhong1995} 
This method follows the spirit of the first-principles-based calculations
described above, but now the model is expanded to allow for different
distortions in each individual unit cell.  The relevant degrees
of freedom are taken to consist of a ferroelectric local-mode vector
in each cell; a displacement vector in each cell; and the global homogeneous
strain variables.  The ferroelectric local-mode and the displacement
variables describe local (in general inhomogeneous) polarizations and
strains, respectively.  The number of parameters needed to describe
such a model is larger, since one needs to include new coefficients that
quantify the interactions between local-mode variables on neighboring sites.
However, all of the coefficients in the expanded model are again obtained
from fitting to an appropriately constructed database of first-principles
results.  Finally, the effective-Hamiltonian model obtained in this
way can be used as the basis for finite-temperature Monte Carlo (MC)
simulations that can be used to map out finite-temperature phase
diagrams.

\begin{figure}
\centerline{\epsfxsize=8cm \epsfbox{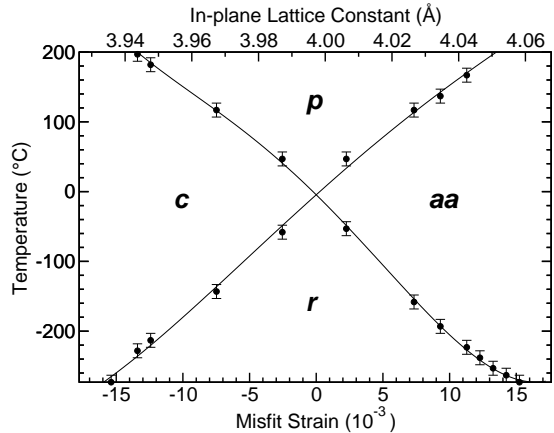}}
\caption{Phase diagram of epitaxial BaTiO$_3$ obtained using the effective 
         Hamiltonian of Zhong, Vanderbilt and Rabe
         \protect\cite{Zhong1994,Zhong1995}. All transitions are second-order.}
\label{fig_diagramHEFF}
\end{figure}

It is straightforward to impose the constraint of fixed in-plane strain by
fixing three of the six elements of the homogeneous strain tensor
during the MC simulations.  For each value of in-plane strain, MC
thermal averages are obtained for the unconstrained components of the
homogeneous strain and the average polarization,
and phase transitions are identified by monitoring the symmetries of these
quantities.  The resulting Pertsev diagram is shown in
Fig.~\ref{fig_diagramHEFF}; it is reproduced from Ref.~\cite{Dieguez2004},
where additional technical details were given.

The Pertsev diagrams of Figs.~\ref{fig_diagramsPERTSEV}(a),
\ref{fig_diagramsPERTSEV}(b), and \ref{fig_diagramHEFF} share the
same topology above and just below $T_{\rm C}$: {\em p} at high
temperature, {\em c} at large compressive misfit, {\em aa} at 
large tensile misfit, and
a four-phase point connecting these phases with the {\em r} phase at
$T_{\rm C}$.  At lower temperature, there is a drastic difference
between Figs.~\ref{fig_diagramsPERTSEV}(a) and
\ref{fig_diagramsPERTSEV}(b), with our theory supporting the
latter.  While our theory underestimates the temperature of
the 4-phase crossing point in Fig.~\ref{fig_diagramHEFF} by about
100~$^\circ$C, this is the price we pay for insisting on a
first-principles approach; indeed, this effective Hamiltonian
underestimates the temperature of the bulk cubic-to-tetragonal
transition by about the same amount.


\section{Studies of other perovskite films}
\label{sec_others}

In this section we review work on the effect of misfit strain
on perovskite films other than BaTiO$_3$ using first-principles 
methodologies similar to the ones described in the previous section.

\subsection{KNbO$_3$}

KNbO$_3$ is a perovskite whose bulk form is very similar to that of
BaTiO$_3$.
It displays the same sequence of transitions with decreasing temperature:
cubic to tetragonal to orthorhombic to rhombohedral, with five-atom
unit cells.

Using our zero-temperature first-principles-base approach, we found
\cite{Dieguez2005} that KNbO$_3$ and BaTiO$_3$ also behave in a very
similar way in terms of their behavior as a function of epitaxial
strain in the context of film geometries.
KNbO$_3$ shows the same $c \rightarrow r \rightarrow aa$ sequence of
transitions as does BaTiO$_3$ when varying the misfit strain at
zero temperature in the absence of external perpendicular stress.
The first transition occurs at a misfit strain of $-4.8 \times 10^{-3}$,
and the second occurs at $5.5 \times 10^{-3}$.

\subsection{PbTiO$_3$}

As in the cases of BaTiO$_3$ and KNbO$_3$, bulk PbTiO$_3$ has a ferroelectric
ground state with a five-atom unit cell. In this case, however, there is
only one transition, from cubic to tetragonal, as the temperature is
reduced.

A very similar approach to the full first-principles one presented
in Sec.~\ref{sec:full} was used by Bungaro and Rabe to study the
influence of misfit strain on PbTiO$_3$ films \cite{Bungaro2004}.
As in the case of BaTiO$_3$, they find a $c \rightarrow r \rightarrow aa$ 
sequence of phase transitions, although in this case the range of stability
of the $r$ phase is narrower than for BaTiO$_3$ (from $+6.6 \times 10^{-3}$
to $+14.8 \times 10^{-3}$), and occurs only at expansive strains.
In the same paper, these authors analyze a superlattice with
alternating layers of PbTiO$_3$ and PbZrO$_3$, finding again the 
$c \rightarrow r \rightarrow aa$ sequence of transitions, but with a wider 
region of $r$ behavior than in the pure PbTiO$_3$ case.

Our first-principles-based approach of Ref.~\cite{Dieguez2005} also predicts
that PbTiO$_3$ adopts the $c$ phase for compressive strains and the $aa$ phase
for tensile strains, with a window of around 1\% misfit strain in between
(from $-3.0 \times 10^{-3}$ to $8.4 \times 10^{-3}$).
However, we find that the $r$ phase is less energetically favorable than
a combination of $c$ and $aa$ domains.

\subsection{SrTiO$_3$}

This perovskite is a ``quantum paraelectric'' in bulk form, meaning that
it is only the zero-point motion of ions that prevents ferroelectricity
from developing.
The cubic phase undergoes a non-ferroelectric oxygen-tilting (or
antiferrodistortive) transition at about 105\,K.

Antons and coworkers \cite{Antons2005} carried out a full first-principles
investigation along the lines we have described in Sec.~\ref{sec:full}
for SrTiO$_3$ films with five atoms in the unit cell.
They found that in this case the sequence of second-order transitions is
$c \rightarrow p \rightarrow aa$ as the strain goes
from compressive to tensile.
The paraelectric phase exists for misfit strains between
$-7.5 \times 10^{-3}$ and $+5.4 \times 10^{-3}$.
They also observed a strong dependence of the dielectric properties of these
films on the misfit strain.
A similar behavior was observed by us in \cite{Dieguez2005}, although with 
a narrower window of $p$ phase in between the $c$ and $aa$ phases.

In their first-principles study, Lin, Huang, and Guo \cite{Lin2006} considered
larger unit cells for SrTiO$_3$, including in this way the possibility
of having antiferrodistortive instabilities in the films.
As in the previous studies, they find that the polarization goes from pointing
in the [001] direction for sufficiently compressive strains to pointing
in the [110] direction for sufficiently tensile strains.
The paraelectric window extends in their case from
$-4 \times 10^{-3}$ to $+4 \times 10^{-3}$.
Their phase diagram is richer due to the presence of the new phases that
involve rotations of the oxygen octahedra.

\subsection{KTaO$_3$}

As in the case of SrTiO$_3$, ferroelectricity is suppressed in bulk KTaO$_3$
by quantum fluctuations.
In Ref.~\cite{Akbarzadeh2007} Akbarzadeh and coworkers use a first-principles
based method to draw the Pertsev diagram for KTaO$_3$ films.
Unlike the theories described above, this work includes electrostatic
boundary conditions and finite-size effects, albeit in an approximate
way.  For films that are 28\,\AA~thick, these authors find that quantum
fluctuations have almost no effect when the electrodes are ideal
(short-circuit boundary conditions).
In this case, the diagram is very similar to that of strong ferroelectrics
like BaTiO$_3$.
When they simulate more realistic electrodes, they find that the corresponding
depolarizing fields in the film strongly affect the phase diagram. 
Under these conditions, the quantum fluctuations might affect the phase
diagram and eventually suppress the $r$ phase.

\subsection{Other perovskites}

Bulk CaTiO$_3$ experiences a phase transition from the cubic phase (with a 
five-atom unit cell) to an orthorhombic phase (with a twenty-atom unit cell)
as temperature is reduced.
Our calculations \cite{Dieguez2005} for CaTiO$_3$ films show a similar
behavior as for the PbTiO$_3$ ones,
with the window between $c$ and $aa$ phases 
(misfit strains of $-2.3 \times 10^{-3}$ to $5.35 \times 10^{-3}$) showing
a mixture of $c$ and $aa$ domains.

The ground state of bulk NaNbO$_3$ shows a ferroelectric monoclinic phase with 
twenty atoms in the unit cell.
The behavior of the film form \cite{Dieguez2005} is similar to that of
KNbO$_3$, with a $c \rightarrow r \rightarrow aa$ sequence of transitions
occurring at misfit strains of $5.5 \times 10^{-3}$ and $4.1 \times 10^{-3}$.

Bulk PbZrO$_3$ undergoes a transition from cubic to an antiferroelectric
phase that has forty atoms in the unit cell.
In Ref.~\cite{Dieguez2005} we show that the five-atom unit cell film stays in
the $r$ phase except for large misfit strains that are experimentally 
difficult to achieve.  The behavior of the true ground-state structure
with epitaxial strain has yet to be fully explored.

BaZrO$_3$ maintains the simple cubic structure in bulk form at all
temperatures.  According to our previous study \cite{Dieguez2005},
it also remains paraelectric in film form for all experimentally
relevant misfit strains,


\section{Summary}
\label{sec_summary}

In this article we have reviewed the role that first-principles calculations
have played in understanding the effects of substrate-imposed
misfit strain on epitaxially grown perovskite ferroelectric films.
To do so we have analyzed the case of BaTiO$_3$, complementing our
previous publications on this subject with unpublished data to help
explain in detail how these calculations are done.
In particular, we have added new explanations and figures to clarify how we
carried out a stability analysis to identify precisely the critical
misfit strains at which phase transitions occur, how
we checked for the possible effects of octahedral rotations, and how we 
computed the coefficients of our first-principles based Landau-Devonshire
theory.
We have also reviewed similar studies in the literature for other perovskite
ferroelectric-film materials.

The use of first-principles calculations to understand the effects of misfit
strain on the properties of films is not restricted to the type of materials
discussed here.
For example, Fennie and Rabe \cite{Fennie2006} have suggested that misfit
strain can be used as a mechanism to obtain strong coupling between
ferroelectric and magnetic ordering in EuTiO$_3$.
Even more recently, Ishida and Liebsch \cite{Ishida2007} have found that, 
when grown epitaxially on SrTiO$_3$, LaTiO$_3$ becomes a highly correlated
metal, instead of being a Mott insulator as found in the bulk form.

We have shown that important insights can be gained by isolating the effects of
misfit strain on the properties of ferroelectric films.
In the future, it would be desirable to extend the theory by developing
similar systematic approaches capable of handling other effects that
are relevant for films, including electric boundary conditions,
domain formation, surface and interface effects, and the role of
vacancies and impurities.


\vspace{1cm}
This work was supported in part by ONR Grant No. N00014-05-1-0054.



\end{document}